\documentclass[12pt]{article}

\usepackage{mathtools}
\usepackage{csquotes}

\usepackage[style=authoryear,firstinits=true,backend=bibtex,maxbibnames=99,doi=false,url=false,isbn=false]{biblatex}
\DeclareFieldFormat{pages}{#1}
\renewbibmacro{in:}{%
  \ifentrytype{article}{}{%
  \printtext{\bibstring{in}\intitlepunct}}}

\title{\textbf{\large{Vertical Temperature Profiles at Maximum Entropy Production with a Net Exchange Radiative Formulation}}}

\author{\textsc{Corentin Herbert}
				\thanks{\textit{Corresponding author address:} 
				Corentin Herbert, Laboratoire des Sciences du Climat et de l'Environnement, 
				CEA Saclay, Orme des Merisiers, 91191 Gif-sur-Yvette, France. 
				\newline{E-mail: corentin.herbert@lsce.ipsl.fr}}\\
\centerline{\textit{\footnotesize{Laboratoire des Sciences du Climat et de l'Environnement, CEA Saclay, Orme des Merisiers, Gif-sur-Yvette, France}}}
\and \textit{\footnotesize{SPHYNX/SPEC/DSM, CEA Saclay, Orme des Merisiers, Gif-sur-Yvette, France}}
\and 
\centerline{\textsc{Didier Paillard}}\\
\centerline{\textit{\footnotesize{Laboratoire des Sciences du Climat et de l'Environnement, CEA Saclay, Orme des Merisiers, Gif-sur-Yvette, France}}}
\and 
\centerline{\textsc{B\'ereng\`ere Dubrulle}}\\
\centerline{\textit{\footnotesize{SPHYNX/SPEC/DSM, CEA Saclay, Orme des Merisiers, Gif-sur-Yvette, France}}}
}
\date{}

\newcommand{\myabstract}{Like any fluid heated from below, the atmosphere is subject to vertical instability which triggers convection. Convection occurs on small time and space scales, which makes it a challenging feature to include in climate models. Usually sub-grid parameterizations are required. Here, we develop an alternative view based on a global thermodynamic variational principle. We compute convective flux profiles and temperature profiles at steady-state in an implicit way, by maximizing the associated entropy production rate. Two settings are examined, corresponding respectively to the idealized case of a gray atmosphere, and a realistic case based on a Net Exchange Formulation radiative scheme. In the second case, we are also able to discuss the effect of variations of the atmospheric composition, like a doubling of the carbon dioxide concentration.}

\begin{document}

\maketitle

\begin{center}
\begin{abstract}
	\myabstract
\end{abstract}
\end{center}

\section{Introduction}

Differential heating in the atmosphere drives atmospheric motion of primary importance for climate. On the horizontal direction, the meridional transport of energy effected by the atmosphere (and the ocean) moderates the temperature gradient as compared to the conditions which would prevail if the planet were in local radiative equilibrium. On the vertical dimension, differential heating may trigger convection, which affects the lapse rate on the long term. Since convection occurs sporadically and on relatively small spatial scales, it is necessary to parameterize its effect in global climate models. As this is an important source of error, it would be desirable to investigate methods to compute the mean convective fluxes with the least possible adjustable coefficients. In fact, horizontal heat transport suffers from the same problem: due to the turbulent nature of the transport, subgrid-models are also required. 
In this context, it was suggested by \textcite{Paltridge1975} that one may use a global thermodynamic variational principle to compute the atmospheric heat transport. More precisely, he conjectured that the meridional fluxes of energy may arrange so as to maximize a function, which corresponds to the rate of entropy produced by the meridional transport, subject to a global constraint. This hypothesis was criticized, rightly, for a number of reasons. First and foremost, there is no theoretical basis for the maximum entropy production (MEP) principle --- see \textcite{Ozawa2003,kleidonlorenzbook,Martyushev2006,DewarBook} for general introductions. Some attempts were made to provide one, notably by \textcite{Dewar2003,Dewar2005}, but as of today a convincing proof is still lacking (see for instance the comments by \textcite{Grinstein2007}), although connections exist with the phenomenological, well-established, theory of non-equilibrium thermodynamics \parencite{DeGrootMazurBook}, and in particular with the Onsager theory which applies close to equilibrium \parencite{Onsager1931}. In addition, the Paltridge model suffers from a number of ad-hoc hypothesis. In spite of this, his results appear to be relatively robust \parencite{Grassl1981,Gerard1990,OBrien1995,Herbert2011b}. 
In the case of the meridional heat transport, even though it may seem counter-intuitive to be able to reproduce a reasonable temperature distribution with no explicit representation of the dynamics of the atmosphere and ocean, one may argue that finding the correct shape for the meridional transport is not so surprising as it can be proved by other means that this feature is constrained by global thermodynamic properties \parencite{Stone1978}. 
On the other hand, we know that heat transport on the vertical involves different physical processes, characterized by faster time scales and smaller spatial scales. In particular, the hydrological cycle \parencite{Pauluis2002a,Pauluis2002b,Pauluis2011} and baroclinicity \parencite{Lucarini2011c} play an important part in the thermodynamics of climate on the vertical dimension.
Nevertheless, it seems appealing to try to apply the maximum entropy production principle to compute vertical heat fluxes without making use of explicit sub-grid parameterizations. Along these lines, \textcite{Ozawa1997} obtained realistic temperature and convective flux profiles using a very simple model for the vertical structure of the atmosphere. In particular, they assumed that the atmosphere absorbed radiation as a gray body. The advantage of this unrealistic assumption is that radiative computations are straightforward. These results were essentially confirmed by further indirect studies in similarly idealized frameworks of a gray \parencite{Pujol2002} or semigray \parencite{Pujol2003} atmosphere. Hitherto, the only study considering the full, non-gray character of atmospheric radiation focused on the role of water-vapor and clouds, treated with the MEP principle, while the thermal structure of the atmosphere was modeled with standard energy-balance methods \parencite{Wang2008}.
On the other hand, sensitivity experiments in a General Circulation Model (GCM) with respect to coefficients involved in sub-grid parameterizations, in which the entropy production rate was considered as a diagnostic quantity, did not support these results \parencite{Pascale2011}. Similarly, the results of a 2D zonal-mean variational problem in which the diabatic heating field is extracted from a GCM show less agreement on the vertical dimension than on the horizontal \parencite{Pascale2012}.
In this study, we propose to compute MEP temperature and convective flux profiles for a realistic atmosphere, in the context of a predictive model formulated as a variational problem. To this end, we have developed a new radiative model, based on the \emph{Net Exchange Formulation} for radiation \parencite{Dufresne2005}. This model is a narrow band model based on the Goody statistical model, taking into account absorption by water vapor and carbon dioxide. We have solved numerically the variational problem for various standard water-vapor profiles. These solutions constitute the first MEP vertical profiles obtained with a realistic radiative scheme. In particular, they do not exhibit the vertical instability that appeared in previous studies with gray atmospheres \parencite{Ozawa1997}. The model also allows for further investigations, like the behaviour of the vertical profiles when the radiative properties of the atmosphere vary. Specific attention is devoted here to the case of a variation in the carbon dioxide concentration.

The outline of the paper is as follows: after briefly presenting MEP profiles in the case of a gray atmosphere in section \ref{graymodelsection}, we show in section \ref{nefprofilessection}  the profiles obtained with a new, realistic radiative scheme, based on the Net Exchange Formulation (NEF). In section \ref{climsensection}, preliminary results about the climate sensitivity of the NEF-based MEP model are shown, while the conclusions are presented in section \ref{conclusection}.

\section{MEP Profiles for a gray atmosphere}\label{graymodelsection}

Let us first briefly treat the idealized case of a gray atmosphere, following \textcite{Ozawa1997}. We consider an atmospheric column made of $N$ layers. Each box has a temperature $T_i$. At the interface between layer $i$ and $i+1$, we note $F_S(i)$ the net incoming (downward) solar flux, $F_L(i)$ the net outgoing (upward) longwave radiation, and $F_C(i)$ the net (upward) convective flux. At steady-state we must have at each interface:
\begin{align}
F_S(i)=F_L(i)+F_C(i).
\end{align}

\subsection{A gray radiative model}

The radiative fluxes $F_S$ and $F_L$ depend on the composition of the atmosphere and on the temperature profile. 

\subsubsection{Shortwave Radiation}
Let us assume here that the atmosphere is purely absorbing in the shortwave domain (no diffusion) and that the absorption is continuous, so that the solar fluxes satisfy the Beer-Lambert law:
\begin{align}
dF_S(\tau)=-F_S(\tau)d\tau,
\end{align}
where $\tau$ denotes the (shortwave) optical depth. We choose atmospheric levels of equal shortwave optical depth, with total optical depth $\tau_S$, so that 
\begin{align}\label{solarfluxeseq}
F_S(i)=F_S(N)e^{-\left(1-\frac i N\right)\tau_S}.
\end{align}
The top-of-the-atmosphere net solar flux is given by $F_S(N)=(1-\alpha_p)S/4$, where $S=1368$ W.m$^{-2}$ is the solar constant and $\alpha_p = 0.3$ the planetary albedo. The total shortwave optical depth is chosen so that the net surface solar flux is $F_S(0) = 142$ W.m$^{-2}$, which gives $\tau_S \approx 0.524812$.

\subsubsection{Longwave Radiation}
For the longwave fluxes, we also assume that the atmosphere is purely absorbing (no diffusion), with a linear extinction, in the framework of the Eddington approximation (two stream aproximation \parencite{LiouBook}). The radiative transfer equation reads
\begin{align}
4\sigma \frac {dT^4}{d\tau}=3F_L(\tau)-\frac{d^2F_L(\tau)}{d\tau^2},
\end{align}
where $\tau$ denotes the longwave optical depth. For atmospheric levels with equal longwave optical depth, the radiative transfer equation can be discretized as
\begin{align}
\sigma T_{i+1}^4-\sigma T_i^4=-\frac {N} {4\tau_L} \left( \left[ 3 \left(\frac {\tau_L}{N}\right)^2+2\right] F_L(i)-F_L(i-1)-F_L(i+1)\right),
\end{align}
where $\tau_L$ is the total longwave optical depth, $\sigma$ the Stefan-Boltzmann constant, and $1\leq i \leq N-1$. Besides,
\begin{align}
\sigma T_1^4-\sigma T_0^4 &= \frac N {4\tau_L} \left( F_L(1)-F_L(0)\right)-\frac {F_L(0)} 2,\\
\sigma T_N^4 &= \frac {F_L(N)} 2 - \frac {N} {4\tau_L} \left( F_L(N-1)-F_L(N)\right).
\end{align}
These equations can be recast in a simple matrix form. Let $\alpha=N/2\tau_L$. We define the $N+1$ dimensional square matrices $\mathbf{A}$ and $\mathbf{S}$ and the $N+1$ dimensional vectors $\mathbf{T}$ and $\mathbf{F}$:
\begin{align}
\mathbf{A} &= 
\begin{pmatrix} 
1 & -1 & 0 & \cdots & 0\\
0 & 1 & -1 & \cdots & 0\\
\vdots & \ddots & \ddots & \ddots & \vdots\\
0 & \cdots & 0 & 1 & -1\\
0 & \cdots & 0 & 0 & 1
\end{pmatrix}, \mathbf{T} = \begin{pmatrix} \sigma T_0^4 \\ \vdots \\ \sigma T_N^4 \end{pmatrix},\\
\mathbf{S} &= 
\begin{pmatrix} 
\frac{\alpha+1} 2 & -\frac {\alpha} 2 & 0 & \cdots & 0\\
-\frac {\alpha} 2 & \alpha+ \frac 3 {8\alpha}  & -\frac {\alpha} 2 & \cdots & 0\\
\vdots & \ddots & \ddots & \ddots & \vdots\\
0 & \cdots & -\frac {\alpha} 2 & \alpha+ \frac 3 {8\alpha} & -\frac {\alpha} 2\\
0 & \cdots & 0 & -\frac {\alpha} 2 & \frac{\alpha+1} 2
\end{pmatrix},
\mathbf{F} = \begin{pmatrix} F_L(0) \\ \vdots \\ F_L(N) \end{pmatrix}.
\end{align}
The radiative transfer equations now simply read 
\begin{align}\label{longwavefluxeseq}
\mathbf{AT=SF}.
\end{align}
The matrices $\mathbf{A}$ and $\mathbf{S}$ being clearly invertible, one may at leisure express $\mathbf{T}$ as a function of $\mathbf{F}$ and conversely.

We further assume that the shortwave and longwave optical depths are proportional, so that we can choose atmospheric levels which have at the same time equal shortwave optical depth and equal longwave optical depth. Then Eqs. \ref{solarfluxeseq} and \ref{longwavefluxeseq} hold simultaneously.

\subsection{The MEP variational problem for convection}

The thermodynamic equation reads
\begin{align}
c_p \frac{\partial T}{\partial t} &= \nabla . \left(F_S-F_L-F_C\right),\\
\shortintertext{or in the discrete form:}
c_p \frac {\partial T_i} {\partial t}&= F_S(i)-F_S(i-1)+F_L(i-1)-F_L(i)+F_C(i-1)-F_C(i).
\end{align}
At steady-state, this is equivalent to $F_S(i)=F_L(i)+F_C(i)$ for $1 \leq i \leq N$.
In these equations, the convective flux $F_C$ is not determined. One possibility would be to build a sub-grid parameterization to compute it. Alternatively, we can formulate the \emph{Maximum Entropy Production} variational problem, which states that, at steady-state, we should choose the values of $F_C(1),\ldots,F_C(N)$ maximizing the entropy production associated to convection. The (thermodynamic) entropy production associated to convection reads
\begin{align}
\sigma &= \sum_{i=0}^N \frac {F_C(i-1)-F_C(i)}{T_i},\\
&=\sum_{i=0}^{N} F_C(i) \left( \frac 1 {T_{i+1}} - \frac 1 {T_i} \right)\label{epdefeq},
\end{align}
where $F_C(-1)=0$. Mathematically, $\sigma$ can be seen either as a function of the $N+1$ variables $F_C(0),\ldots,F_C(N)$, because the temperatures can be expressed as $\mathbf{T=A^{-1}SF}$ and the longwave fluxes $\mathbf{F}$ are themselves functions of the convective fluxes at steady-state through the relation $F_L(i)=F_S(i)-F_C(i)$, or as a function of the $N+1$ variables $T_0,T_1,\ldots,T_N$ by reversing these relations. Let us for instance see $\sigma$ as a function of the convective fluxes, so that the MEP variational problem simply reads:
\begin{align}
\max_{(F_C(0),\ldots,F_C(N))} \{ \sigma(F_C(0),\ldots,F_C(N)), F_C(N)=0 \},
\end{align}
where the global constraint $F_C(N)=0$ enforces the steady-state condition. This is a constrained variational problem, but in practice it can be solved easily by deleting the term corresponding to $i=N$ in the expression for the entropy production (\ref{epdefeq}) and in the $\mathbf{F}$ vector.
Alternatively, considering $\sigma$ as a function of the temperatures, the variational problem reads
\begin{align}\label{grayvarprobeq}
\max_{(T_0,\ldots,T_N)} \sigma(T_0,\ldots,T_N).
\end{align}
This is an unconstrained variational principle (the constraint $F_C(N)=0$  can again be enforced simply by deleting the term corresponding to $i=N$ in the expression for the entropy production (\ref{epdefeq})), and as such it is easy to solve numerically. Contrary to the two-dimensional horizontal case \parencite{Herbert2011b}, the treatment of the steady-state constraint is thus straightforward here. The reason for this is that the formulation of the problem as a one-dimensional model with fluxes at the cell boundaries reduces the condition --- vanishing integral of the divergence --- to a condition on a single variable.

\subsection{Results for a gray atmosphere}

\begin{figure*}[t]
\noindent\includegraphics[width=0.5\textwidth]{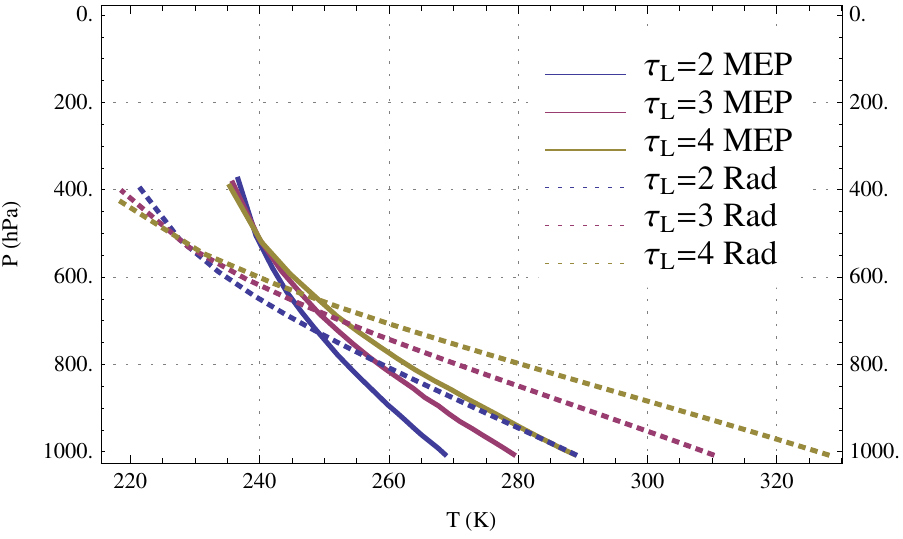}
\includegraphics[width=0.5\textwidth]{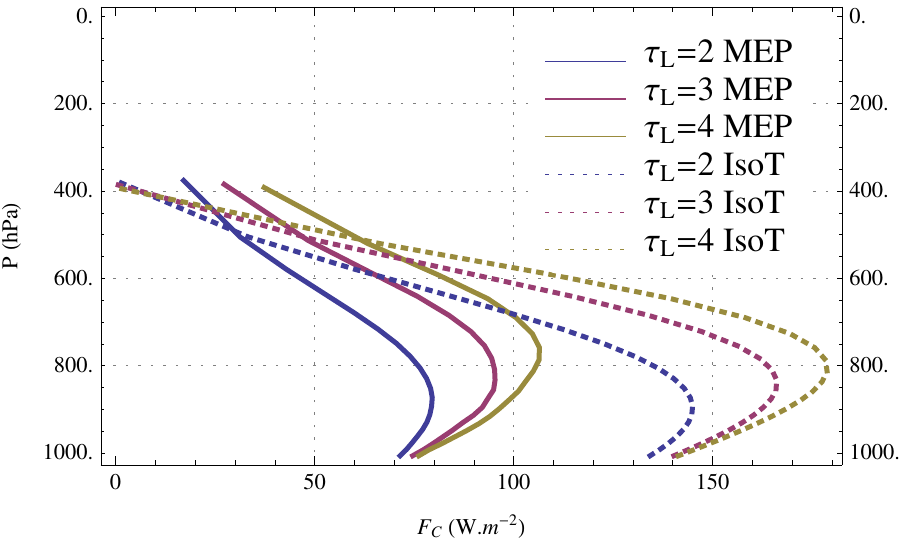}
\caption{Left: Temperature profiles at MEP state (solid lines) and at radiative equilibrium (dashed lines) for various longwave optical depths: $\tau_L=2$ (blue), $\tau_L=3$ (red) and $\tau_L=4$ (yellow). Right: Convective flux profiles at MEP state (solid lines) and for isothermal profiles (dashed lines) for various longwave optical depths: $\tau_L=2$ (blue), $\tau_L=3$ (red) and $\tau_L=4$ (yellow).}\label{graymeppflsfig}
\end{figure*}

Given a set of numerical values for the parameters of the models, solving the model reduces to solving the variational problem (\ref{grayvarprobeq}). The parameters of the model are the solar constant $S$, the planetary albedo $\alpha_p$, the shortwave optical depth $\tau_S$ and the longwave optical depth $\tau_L$. Their values are recorded in Table \ref{grayparamstable}.
\begin{table}[t]
\caption{Gray atmosphere model parameters.}\label{grayparamstable}
\begin{center}
\begin{tabular}{cccc}
\hline\hline
$S$ & $\alpha_p$ & $\tau_S$ & $\tau_L$ \\
\hline
$1368$ W.m$^{-2}$  & 0.3 & 0.524812 & 2 -- 4\\
\hline
\end{tabular}
\end{center}
\end{table}
The variational problem can be solved using standard optimization routines. A unique maximum is found. The resulting profiles are shown on Fig. \ref{graymeppflsfig} for various choices for the longwave optical depth.
Note that the convective fluxes slightly differ from those obtained by \textcite{Ozawa1997} with the same model: we find an inversion of the convective flux profile in the lower troposphere while they had found a constant profile in this region. This profile does not appear as a maximum entropy production state in our model, not even a local one. A possible interpretation of this behavior is that the entropy production rate is locally negative in the inversion zone for the unconstrained variational problem. This locally negative entropy production rate is compensated for by a larger (positive) entropy production rate in the overlying regions, resulting in an overall entropy production rate higher than for monotonically decreasing flux profiles. Note that a locally negative entropy production rate does not contradict the second law of thermodynamics as it only states that the entropy of the universe should grow, that is, the global --- not local --- entropy production rate should be positive. Nevertheless, imposing a positive local entropy production as an additional constraint in the variational problem, we do recover a profile similar to \textcite{Ozawa1997}.

\begin{figure}[t]
\centering
\includegraphics[width=0.5\textwidth]{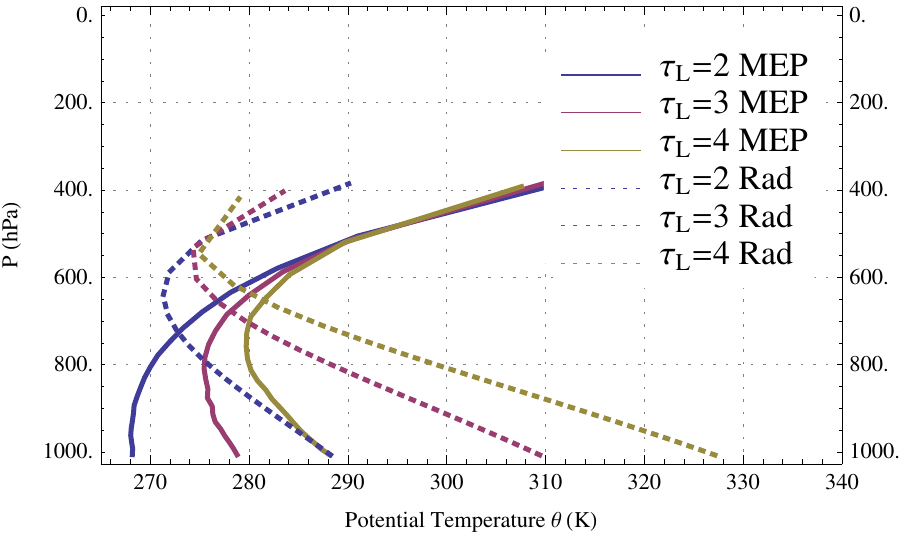}
\caption{Potential Temperature Profiles at MEP state (solid lines) and at radiative equilibrium (dashed lines) for various longwave optical depths: $\tau_L=2$ (blue), $\tau_L=3$ (red) and $\tau_L=4$ (yellow).}\label{graymepthetapflsfig}
\end{figure}

Also represented on Fig. \ref{graymeppflsfig} are the temperature profile obtained at radiative equilibrium (that is, all the convective fluxes set to zero) and the convective flux profile obtained by imposing an isothermal temperature profile. These two cases correspond to the bounding profiles for which the entropy production rate is zero. Compared to the radiative equilibrium profile, the MEP profile has a much more moderate vertical temperature gradient. Similarly, the convective fluxes at MEP state are roughly half those needed to maintain a isothermal temperature profile.
Note that the temperature profiles so obtain are not statically stable: Fig. \ref{graymepthetapflsfig} shows the vertical profiles of (dry) potential temperature, defined by
\begin{align}
\theta &= T \left(\frac{P_0}{P}\right)^{\kappa}, \quad \kappa=\frac{R_d}{c_p},
\end{align}
for both the radiative equilibrium profiles and the MEP profiles. As expected, the radiative equilibrium profiles are not statically stable (potential temperature is a decreasing function of height up to roughly $600$ hPa). More interestingly, the MEP profiles are not statically stable either, as the potential temperature is a decreasing function of height up to about $800$ hPa --- which corresponds to the negative entropy production rate zone. Note that it could be possible to impose static stability as a constraint in the variational problem, possibly with an outcome comparable to imposing a locally positive entropy production rate. This constraint would be analogous to the method of \emph{convective adjustment} \parencite{Manabe1964}. However, we shall see in the next section that it is not necessary to do so; replacing the gray radiative model by a realistic radiative model turns the MEP profiles into statically stable profiles.

The temperature and convective flux profile obtained at MEP state with this simple gray atmosphere radiative model are qualitatively realistic. Nevertheless, some criticism may still apply. In the first place, the validity of the MEP hypothesis hitherto lies only on empirical evidence. Besides, the radiative hypothesis are highly idealized. Diagnostic studies of the entropy production rate and other thermodynamic quantities with complex models \parencite{Pascale2011} do not  fully agree with the MEP hypothesis on the vertical dimension. In particular, many different processes contribute to the entropy budget \parencite{Pauluis2002a,Pauluis2002b} and one can think of two atmospheres with the same total entropy production rate but radically different processes responsible for this entropy production \parencite{Volk2010}. However, it is not clear how limited the relevance of diagnostic studies with complex models for assessing the validity of the MEP hypothesis can be, insofar as the formulation of the variational problem implies some degree of freedom in the system. In the absence of any \emph{unknown} degree of freedom, that is when all the degrees of freedom are explicitly resolved, possibly through sub-grid parameterizations, there is no room for an optimization principle anyway, as the dynamics of the system is fully determined by the initial data. 
Hence, to further inquire on the relevance of the MEP variational problem for vertical heat fluxes, we suggest to use an intermediate approach by solving the variational problem with a realistic radiative model, but still no explicit representation of convection. This approach is presented in the next section.

\section{MEP Profiles with the NEF model}\label{nefprofilessection}

\subsection{A NEF band model}\label{nefmodelsection}

In this section, we briefly describe a new radiative model based on the \emph{Net Exchange Formulation} \parencite{Dufresne2005}. The purpose of this model is to reach a balance between a realistic description of the absorption properties of the major radiatively active constituents of the terrestrial atmosphere while keeping a relatively smooth dependence of the radiative flux with respect to the temperature profile. This last requirement is necessary to use the model in the framework of a variational problem. The model presented briefly in this section is described in more details in the online supplementary material.

The fundamental idea of the \emph{Net Exchange Formulation} \parencite{Green1967} is to replace the traditional description of energy exchanges in terms of fluxes by a description in terms of rates of energy exchange between two given layers. In the flux formulation, the description is local: for a given level in the atmosphere, we are interested in the upward and downward fluxes at this precise level. The local laws of emission and absorption allow one to write the radiative transfer equation, the solutions of which give the profiles of radiative fluxes. On the contrary, in the Net Exchange Formulation, we consider two layers in the atmosphere --- say $i$ and $j$ --- and, using the same laws of emission and absorption as usual, we write the net energy exchange rate (NER) between these two layers at wavelength $\nu$, denoted $\psi_{ij}^{\nu}$:
\begin{align}
\psi_{ij}^\nu = \int_{\Sigma_i}  dP_i \int_{\Sigma_j}dP_j \int_{\Gamma(P_i,P_j)} d\gamma \alpha_\nu(P_i,\gamma) \alpha_\nu(P_j,\gamma)(B_\nu(T_j)-B_\nu(T_i))I_\nu(\gamma),
\end{align}
where $\Gamma(P_i,P_j)$ denotes the set of all optical paths between point $P_i$ of layer $i$ and point $P_j$ of layer $j$, $\Sigma_i$ the volume of layer $i$ (or a boundary surface --- the surface of the Earth or space), $\alpha_\nu(P_i,\gamma)$ is the absorption coefficient (1 if $i$ is a boundary surface rather than a layer), $B_\nu(T)$ the Planck function, $\displaystyle I_\nu(\gamma) = \exp \left( - \int_\gamma k_\nu(s)ds\right)$, and $k_{\nu}$ is the extinction coefficient. The NER satisfy simple elementary properties, like antisymmetry: $\psi_{ij}^\nu=-\psi_{ji}^\nu$, energy conservation $\sum_{i,j} \psi_{ij}^\nu=0$ (this is a consequence of antisymmetry) or the second law of thermodynamics ($\psi_{ij}^{\nu}$ has the same sign as $T_j-T_i$). Here we are only interested in the spectrally integrated NER: $\displaystyle \psi_{ij}=\int_0^{+\infty}\psi_{ij}^\nu d\nu$.

In the longwave domain, we decompose the spectrum in 22 narrow bands, and in each band we account for absorption by water vapor and carbon dioxide only. The absorption coefficient is computed by using the \textcite{Goody1952} statistical model with the data from \textcite{Rodgers1966}. For the spatial integration, the diffusive approximation is performed with the standard diffusion factor $\mu \approx 1/1.66$.
Apart from the absorption data, given once and for all, the inputs of the model are the water vapor density and temperature profiles and carbon dioxide concentration. Of course, one may either fix absolute or relative humidity.

In the shortwave domain, absorption by water vapor and ozone is accounted for by adapting the parameterization from \textcite{Lacis1974}. Apart from the absorption functions determined thanks to laboratory data, the input parameters for the model are the water-vapor density and ozone density profiles, as well as surface albedo and solar constant.

The radiative budget of an atmospheric layer is given simply by summing over all the NER terms involving the layer in question: for layer $i$, the radiative budget $R_i$ is given by
\begin{align}
R_i = \sum_{j=0}^{N+1} \psi_{ij}
\end{align}
This is quite different from the usual flux formulation for which the radiative budget is the divergence of the flux. In particular, it allows for finer discussions as one may decompose the radiative budget into the NER matrix $\psi_{ij}$ to see which contributions dominate in the radiative exchanges.

\subsection{The variational problem}

In this section,we consider again a one-dimensional atmospheric column (with no lateral fluxes), except that contrary to section \ref{graymodelsection}, the radiative budget is now computed using the NEF model presented in section \ref{nefmodelsection}. The radiative budget is no longer expressed as the divergence of a flux as previously, but rather directly in terms of the net rate of energy exchanged with all the other layers: let us write $R_i(T_0,T_1,\ldots,T_N)$ for the radiative budget of layer $i$ and $\gamma_i$ the convergence of the convective flux (hence $\gamma_i=F_C(i-1)-F_C(i)$). The thermodynamic equation reads:
\begin{align}
c_p \frac{\partial T_i}{\partial t} = R_i(T_0,\ldots,T_N)+\gamma_i.
\end{align}
At steady-state, we simply have $R_i(T_0,\ldots,T_N)+\gamma_i=0$ in each layer ($1 \leq i \leq N$), but $\gamma_i$ is not a known function of $T_i$. The rate of thermodynamic entropy production associated to convection is still given by
\begin{align}
\sigma(T_0,\ldots,T_N) &= \sum_{k=0}^N \frac{\gamma_i}{T_i},\\
&=- \sum_{k=0}^N \frac{R_i(T_0,\ldots,T_N)}{T_i},\\
\shortintertext{with the global constraint}
\sum_{k=0}^N \gamma_k &= \sum_{k=0}^N R_k(T_0,\ldots,,T_N) =0.
\end{align}
Thus, the MEP variational problem reads
\begin{align}\label{nefvarprobeq}
\max_{(T_0,\ldots,T_N)} \left\{ \sigma(T_0,\ldots,T_N)\quad \Big| \quad \sum_{k=0}^N R_k(T_0,\ldots,,T_N) =0 \right\}.
\end{align}
This variational problem is equivalent to Eq. \ref{grayvarprobeq} except that here the radiative budget is computed with a different radiative model. As a consequence it is now more convenient to work with the convergence of convective flux rather than the fluxes themselves, which has the consequence that a constraint appears (this constraint is equivalent to $F_C(N)=0$). Furthermore, writing an equivalent variational problem in terms of the $\gamma_i$ variables would imply to invert the radiative budget. Unlike the case of section \ref{graymodelsection}, it is not straightforward to do so here, because of a more realistic formulation of the radiative exchanges. Note that the advantage of formulating the problem in terms of the energy convergence $\gamma$ is that it makes clear that the structure of the problem is similar to the case of horizontal heat transport \parencite{Herbert2011a,Herbert2011b}. In particular, the variational problem written in this form would remain valid if we considered lateral fluxes as well.

\subsection{Results for the NEF model}

Solving the variational problem (\ref{nefvarprobeq}) with standard optimization algorithms, we observe that there is a unique solution for a given set of parameters. The input parameters for the models are the number of atmospheric layers, the water-vapor density profile (fixing either the absolute humidity or the relative humidity), the ozone profile, the carbon dioxide concentration and the surface albedo (see Table \ref{nefparamstable}).
As the number of vertical layers $N$ increases, the MEP temperature and convective flux profiles converge. A good approximation is already reached for $N=9$ and we shall keep this setting for the rest of this paper. For the other parameters, we choose classical values; we use several standard water vapor density profiles \parencite{McClatchey1972}, and we keep the pre-industrial CO$_2$ concentration in this section.
\begin{table}
\caption{Input Parameters for the NEF Radiative Model:}\label{nefparamstable}
\begin{center}
\begin{tabular}{ccc}
\hline
Parameter & Symbol & Value\\
\hline
\hline
Vertical Resolution & $N_{lev}$ & $9$\\
Diffusion factor & $\mu$ &  $1/1.66$\\
\begin{tabular}{c}Water vapor profile \\ Ozone profile \end{tabular} & $\begin{array}{c}\rho_{H_20}(p)\\ \rho_{O_3}(p)\end{array}$ &  $\left\lbrace\begin{tabular}{l}Trop \\ SAS \\ SAW \\ MLS \\ MLW \end{tabular}\right.$\\
Carbon Dioxide concentration & $\rho_{CO_2}$ & 280 ppmv\\
Solar Constant & $S$ & 1368 W.m$^{-2}$\\
Surface Albedo & $\alpha_g$ & $0.1$\\
\hline
\end{tabular}
\end{center}
\end{table}

\begin{figure*}
\includegraphics[width=0.5\textwidth]{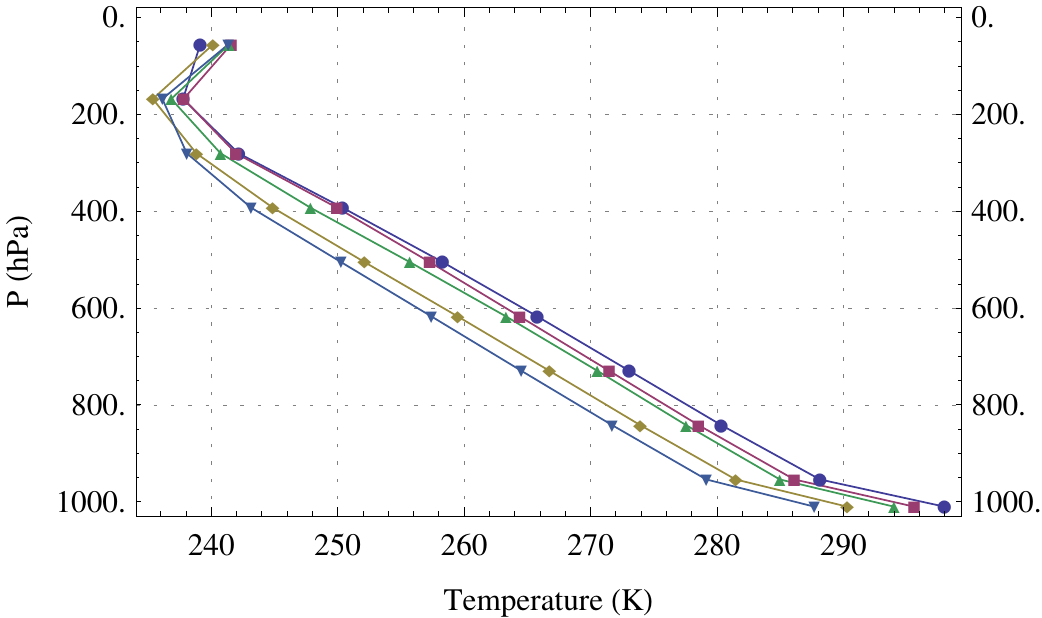}
\includegraphics[width=0.5\textwidth]{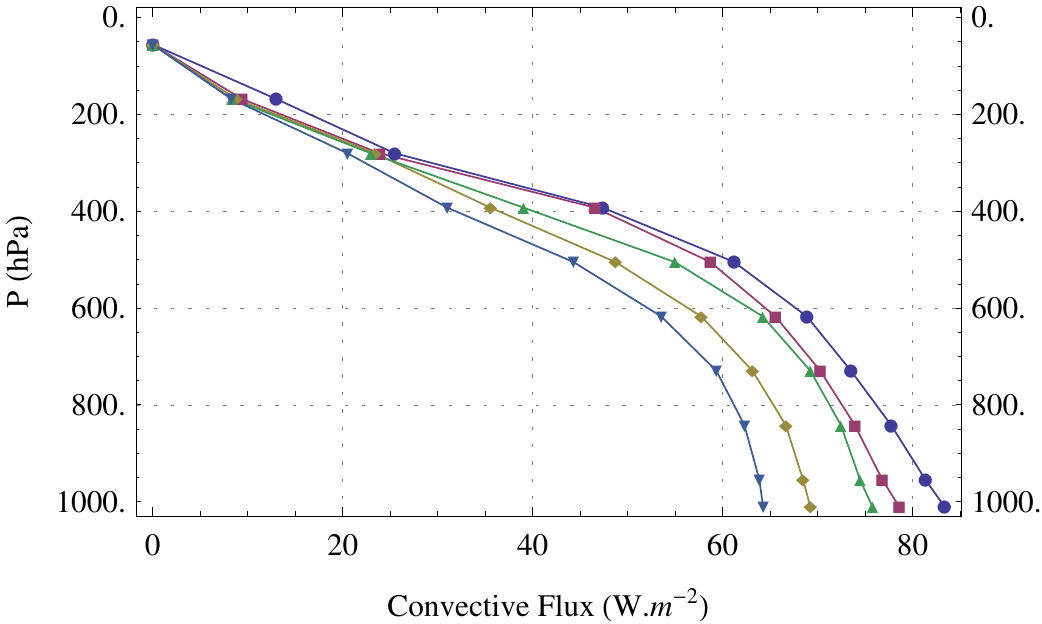}
\caption{Left: MEP temperature profiles for the standard atmospheric profiles: Tropical (blue), Mid-Latitude Summer (red), Mid-Latitude Winter (yellow), Sub-Arctic Summer (green) and Sub-Arctic Winter (light blue). Right: MEP convective flux profiles for the standard atmospheric profiles: Tropical (blue), Mid-Latitude Summer (red), Mid-Latitude Winter (yellow), Sub-Arctic Summer (green) and Sub-Arctic Winter (light blue).}\label{nefmeppflsfig}
\end{figure*}

\begin{figure}[t]
\centering
\includegraphics[width=0.5\textwidth]{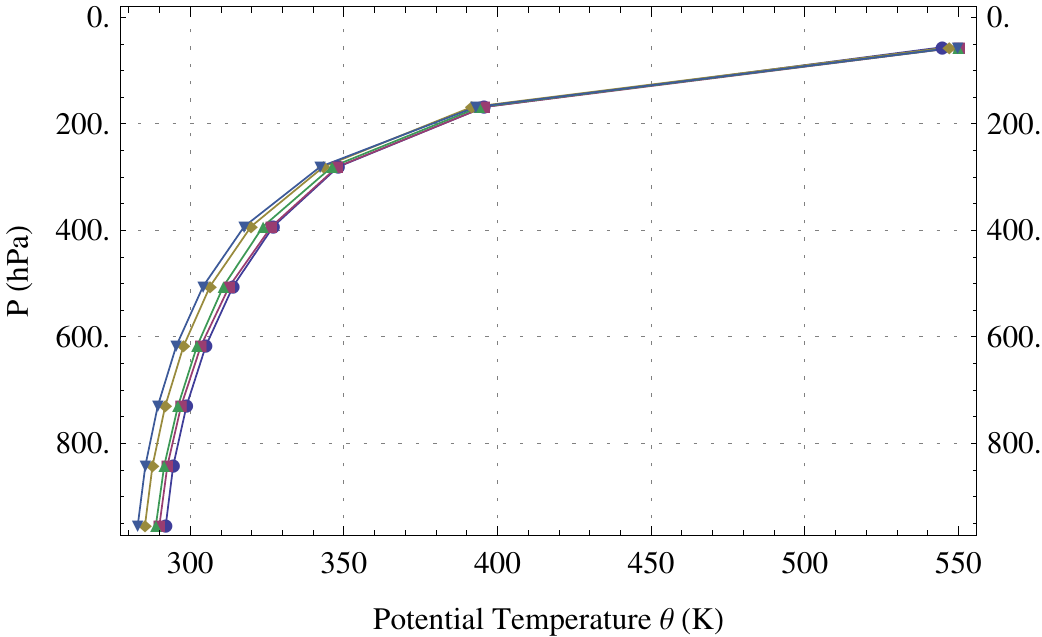}
\caption{Potential temperature profiles at MEP state for the standard atmospheric profiles: Tropical (blue), Mid-Latitude Summer (red), Mid-Latitude Winter (yellow), Sub-Arctic Summer (green) and Sub-Arctic Winter (light blue).}\label{nefthetameppflsfig}
\end{figure}

The temperature and convective flux profiles obtained at MEP state for the standard profiles (fixing the absolute humidity) are shown on Fig. \ref{nefmeppflsfig}.
Compared to the gray atmosphere, we find no inversion in the convection profile, that is no region with a locally negative entropy production. The vertical temperature gradient is also less steep than for the gray atmosphere. Note that the vertical temperature gradient is not very sensitive to the water-vapor content of the atmosphere: for the various standard profiles, the temperature profiles are only shifted towards warmer temperatures when the water-vapor content increases.
Beyond the general structure of the temperature and convective flux profiles, one may wonder if the MEP profiles are statically stable. To this end, we define the (dry) potential temperature as in section \ref{graymodelsection} by
\begin{align}
\theta &= T \left(\frac{P_0}{P}\right)^{\kappa}, \quad \kappa=\frac{R_d}{c_p},
\end{align}
and compute the potential temperature profiles corresponding to the MEP profiles discussed above, represented on Fig. \ref{nefthetameppflsfig}.
In all cases, $\theta$ decreases with altitude, which means that the profiles are statically stable. On the contrary, the MEP profiles obtained for the gray atmosphere in section \ref{graymodelsection} were not statically stable.
It is worth emphasizing that statically stable temperature profiles arise here as solutions of the MEP variational problem without specifying this as a dynamical constraint. This means that this basic property of steady-state profiles is well reproduced by the MEP principle, provided that the radiative model is sufficiently realistic. In particular, there is no need here to patch the MEP model with a convective adjustment as \textcite{Pujol2002} did for the case of a gray atmosphere.
The surface temperature obtained at MEP state for the various standard atmospheric profiles is shown in Table \ref{surftemptable}. The results for the tropical and mid-latitude summer profiles are quite reasonable, but the surface temperature for the mid-latitude winter and sub-arctic profiles seems relatively high. Two factors can account for this high temperature at MEP state: first of all, the effect of clouds is not taken into account, and second, the results presented here are all obtained with a fixed value for the surface albedo: $\alpha_g=0.1$. Hence the idealized conditions we are using here with the idealized profiles to investigate the structure of the MEP profiles is not necessarily meant to represent as closely as possible realistic cases. In particular, the arctic profiles are interesting in that they feature a lower content of water vapor, rather than because they correspond to some geographical location.

\begin{table*}
\caption{Surface temperature and climate sensitivity (surface temperature variation for a CO$_2$ doubling experiment) for different water-vapor profiles, at MEP state, for the NEF radiative model:}\label{surftemptable}
\begin{center}
\begin{tabular}{cccc}
\hline
& MEP Surface Temperature & MEP Climate Sensitivity \\
\hline
\hline
Tropical 				& 24.8 & 0.66 \\
Mid-Latitude Summer 	& 22.4 & 0.60 \\
Mid-Latitude Winter 		& 17.1 & 0.36 \\
Sub-Arctic Summer 		& 20.8 & 0.50 \\
Sub-Arctic Winter 		& 14.6 & 0.24 \\
\hline
\hline
\textcite{Dufresne2008}		& ---	   & 0.4 $\pm$ 0.3\\
\hline
\end{tabular}
\end{center}
\end{table*}

\section{Variations in atmospheric composition}\label{climsensection}

In the previous section, we have shown that replacing the ideal assumptions for radiation used in section \ref{graymodelsection} by a realistic radiative model yields reasonable temperature and convective flux profiles. An advantage of this new radiative scheme is that it allows for variations of the physical parameters characterizing the absorption properties of the atmosphere. Of particular interest are the variations of the concentration of carbon dioxide in the atmosphere. As a test bed, we will consider in this section the effect on the temperature profiles of a doubling of the carbon dioxide concentration. This is a classical experiment to which virtually all the climate models are subjected \parencite{ipccAR4models}. One should however remain aware that with the present MEP model, very little feedback processes are included, while these feedbacks --- for instance the water-vapor, albedo, lapse rate or clouds feedbacks --- usually play a big role in amplifying the climate sensitivity (see e.g. \textcite{Schlesinger1987,Meehl2004,Bony2006,Soden2006}).

\begin{figure}
\centering
\includegraphics[width=0.5\textwidth]{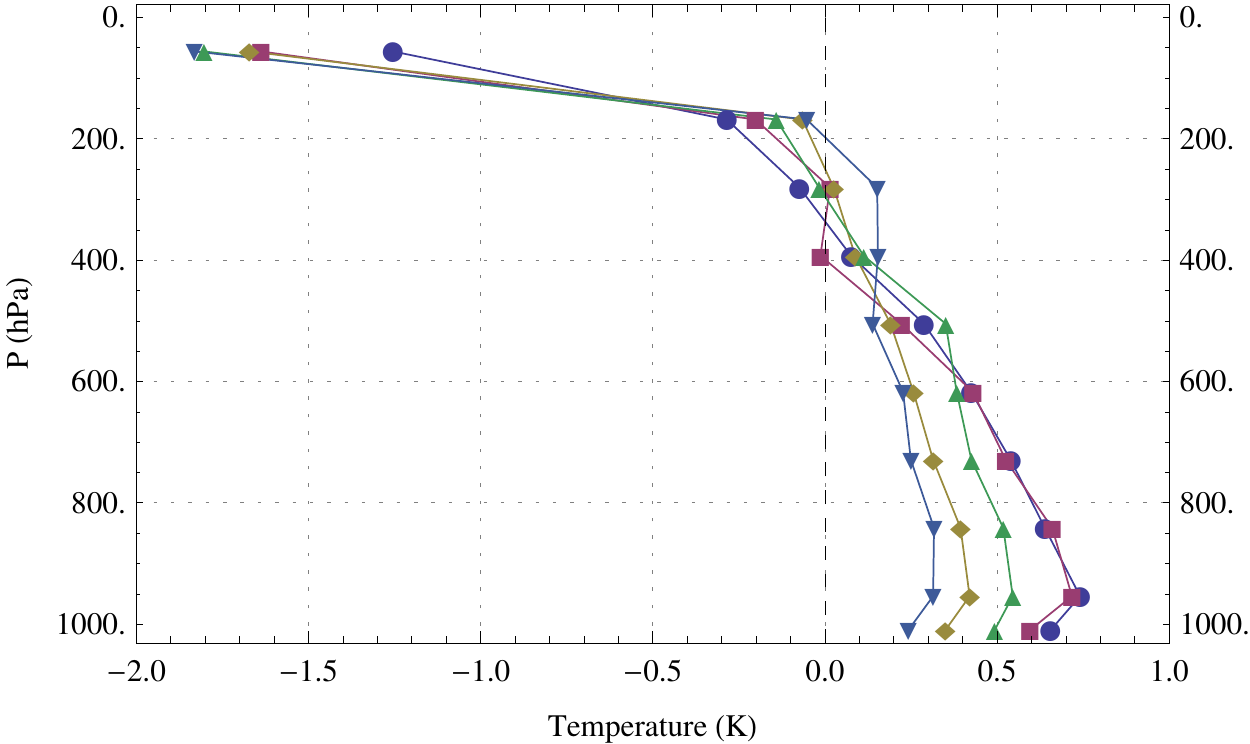} 
\caption{Temperature response for a CO$_2$ doubling experiment for the standard atmospheric (water-vapor) profiles: Tropical (blue), Mid-Latitude Summer (red), Mid-Latitude Winter (yellow), Sub-Arctic Summer (green) and Sub-Arctic Winter (light blue).}\label{nefco2doublingpflfig}
\end{figure}

Figure \ref{nefco2doublingpflfig} shows the response of the temperature profiles obtained with the MEP model --- using the NEF radiative scheme --- for a doubling of the CO$_2$ concentration.
As expected, in the lower atmosphere, the temperature increases when the CO$_2$ concentration increases, while the opposite response is observed in the upper atmosphere. The temperature response to CO$_2$ doubling is greater for profiles with a higher absolute humidity.

The response of the surface temperature to the variation of the carbon dioxide concentration --- usually called \emph{climate sensitivity} --- ranges from $ 0.24$ K (for the sub-arctic winter profile) to $ 0.66$ K (for the tropical profile), as shown in table \ref{surftemptable}. 
To compare these values with the literature, we need to be careful about the feedbacks included in the model we wish to compare to. Indeed, if the overall climate sensitivity is still a subject of debate, this is mainly due to poorly understood feedbacks, like the cloud feedback \parencite{Stephens2005}, which are not accounted for in the present study. On the contrary, the \enquote{basic} response of climate to CO$_2$ doubling --- due to a change in the Planck function \parencite{Hansen1984} --- is well agreed-upon: in terms of temperature, a rise of 1.2 K is considered an accurate estimate. The \enquote{historical} estimate of the seminal study by \textcite{Manabe1967} with a radiative model with convective adjustment was 1.3 K. Other early studies with radiative-convective models with no or little feedbacks provided values as low as 0.8 K \parencite{Hummel1981a,Hummel1981b} or 0.7 K \parencite{Hunt1981} --- see also \textcite{Schlesinger1987}. More recently, multi-model means with state-of-the-art models separating the contributions of various feedbacks \parencite{Dufresne2008} revealed an estimate of $ 0.4 \pm 0.3$ K, when taking into account only the effect of the Planck function and the lapse-rate feedback, as relevant here. The values obtained with the MEP model for idealized atmospheric profiles coincide well with this range.
Further studies would be necessary to compare more accurately the climate sensitivity of a MEP model --- possibly including feedbacks like the water-vapor feedback \parencite{Herbert2012} or the albedo feedback \parencite{Herbert2011a} --- to that of radiative-convective models, or even general circulation models.

\section{Conclusion}\label{conclusection}

We have investigated the possibility of computing the vertical transport of energy in the atmosphere using a thermodynamic variational principle. The variational principle relies on the \emph{Maximum Entropy Production} conjecture \parencite{Ozawa2003,kleidonlorenzbook,Martyushev2006,DewarBook}, which assumes that the energy fluxes at steady state coincide with the values maximizing the associated thermodynamic entropy production. In spite of the lack of theoretical understanding for this hypothesis, the MEP conjecture has proven useful in several practical applications, in particular in climate sciences \parencite{Herbert2012}. The advantage of this formulation is that the energy transport is computed in an implicit way, which means that no explicit form is required for the parameterization of the sub-grid phenomena effecting the vertical transport in reality. In particular, the need for coefficient adjustment is  reduced. In addition, it is also a very fast solution, numerically speaking.

In this study, we have tested the realism of the steady-state temperature and convective flux profiles obtained by representing the vertical energy transport  with the MEP principle. Two frameworks have been explored: in a first step, we have discussed the highly idealized case of a gray atmosphere already considered by \textcite{Ozawa1997}. In a second step, we have built a model based on realistic assumptions for radiation. Specifically, we have designed a radiative scheme based on the Net Exchange Formulation \parencite{Dufresne2005} taking into account narrow spectral bands for the absorption by water-vapor and carbon dioxide. For several realistic atmospheric profiles, we have obtained MEP temperature and convective flux profiles with a satisfactory accuracy. We have shown that in this framework, it is possible to obtain small but realistic responses for the temperature profiles when the composition of the atmosphere is modified. In particular, we have described preliminary results for climate sensitivity obtained in a CO$_2$ doubling experiment.

These results show that in spite of what diagnostic studies based on General Circulation Models seem to indicate \parencite{Pascale2011}, the MEP hypothesis gives qualitatively realistic profiles when used in conjunction with a simple but realistic radiative scheme, even in --- and perhaps because of --- the absence of explicit representation for the hydrological cycle or climate feedbacks like clouds, albedo, water-vapor,... Nevertheless, inclusion of such processes in a MEP model would be crucial for quantitative applications, and in particular for more precise estimations of the climate sensitivity. It also paves the way to applications of this type of models to climate problems of direct relevance, by varying --- at very small computational cost---parameters such as carbon dioxide concentration.

\printbibliography

\end{document}